\documentstyle[12pt]{article}
\begin{document}


\begin{center} {\bf INFORMATION LOSS PARADOX TESTED ON CHIRAL FERMION COUPLED
TO A BACKGROUND DILATONIC FIELD }\\
Anisur Rahaman, Durgapur Govt. College, Durgapur - 713214, Burdwan, West Bengal, INDIA\\
e-mail: anisur@.saha.ac.in \end{center}

\vspace{.5cm}
\noindent PACS No. 11.10 Kk, 11.15. -q\\
Keywords Information loss, Chiral Fermion, Dilatonic Gravity

\vspace{2cm}
\begin{center} {\bf Abstract}
\end{center}

A model where chiral boson is coupled to a background dilaton
field is considered to study the s-wave scattering of fermion by a
back ground dilatonic black hole. Unlike the conclusion drawn in
\cite{MIT} it is found that chiral fermion does not violate
unitarity and information remains preserved. It is found that
Faddevian anomaly plays a crucial role on information loss.
\newpage

In recent years there has been a lot of interest in the physics of
information loss. Matter falling into the black holes carries some
information with it. That becomes inaccessible to the rest of the
world. A problem arises when the black hole evaporates through
Hawking radiation. It is a controversial issue whether or not
quantum coherence would be maintained during the formation and
subsequent evaporation of a black hole.
Quite a long ago Hawking suggested that the process did not
preserve information and unitarity failed to be maintained.
\cite{HAW}. It was an indication of a new level of
unpredictability in the realm of quantum mechanics induced by
gravity. There were plenty of opinions that went against Hawking's
suggestion. The main theme of those opinions was that the
information about the initial state of the system was carried by
some Plank scale steady remnant \cite{HOOFT, GARF}. It is also
fair to admit that the issue gradually shifted against Hawking's
suggestion but it was not well settled.

In a recent publication, we find that Hawking has moved away from
his previous belief and suggested that quantum gravity interaction
does not lead to any loss of information. So there will be no
problem to maintain quantum coherence during the formation and
subsequent evaporation of the black hole \cite{HAW1}.
In spite of that, it has been standing as a controversial issue.
Even now Hawking radiation effect on fermion information loss
problem is not well understood \cite{AHN}.

This type of problem is very difficult to analyze in general.
However, there exist some less complicated models and those were
solved to study this paradox \cite{GIDD, GIDD1, STRO, CALL, SUS}.
In these studies only the s-wave scattering of fermion incident on
the extremal charged black hole was considered. Angular momentum
coordinate becomes irrelevant in this situation and a two
dimensional effective action results \cite{HOOFT}. Though those simplified
models do not capture the detailed physics of black hole those
models contain the information loss paradox in a significant
way \cite{STRO, SUS, MIT}.

To study the information paradox several authors studied the
scattering of fermion \cite{GIDD, GIDD1, STRO, CALL}. The
scattering of boson was also considered in \cite{FIS1, FIS2,
FIS3}. In this context, the scattering of chiral fermion off
dilaton black hole is of particular interest \cite{MIT}. The
scattering of Dirac fermion itself is an interesting problem
\cite{GIDD, GIDD1, STRO, CALL, FIS1, FIS2, FIS3}. If Dirac
fermions are replaced by the chiral fermion it makes the analysis
more complex
because the chiral fermions generate anomaly in the energy
momentum sector when they couple to gravity \cite{BEL}. Therefore,
to get a solution for this type of complicated problem from the
study of the s-wave scattering of fermion from a back ground
dilaton field is interesting in its own right.

In ref. \cite{MIT}, the authors showed that the scattering of
chiral fermion can be studied in the presence of the anomalies if
one deals with the bosonized lagrangian. Bosonization allows
anomaly to enter into the process. It would be intersting to 
investigate the change (if any) appear in the presence of 
anomaly or with the alteration of the same.

In \cite{MIT}, the author showed that the chiral fermion does not
preserve information whereas Dirac fermion in \cite{STRO, SUS}
gave a completely opposite result. It preserved information.
Therefore, one might think that there might be a possiblity for
chiral fermion that will be in agreement with the Hawking's recent
suggestion.
In this context, it would certainly be
interesting to investigate whether chiral fermion can offer
information preserving result like Dirac fermion.
This motivates us to investigate the scattering of chiral fermion
off dilatonic black hole in a new setting.

To this end we consider a model where chiral fermion gets coupled
to a background dilaton field $\Phi$. Of course, electromagnetic
background is there. For sufficiently low energy incoming fermion,
the scattering of s-wave fermion incident on a charge dilaton
black hole can be described by the action
\begin{equation}
{\cal S}_f = \int d^2x[i\bar\psi\gamma^\mu[\partial_\mu +
ieA_\mu]\psi - {1\over 4} e^{-2\Phi(x)}F_{\mu\nu}F^{\mu\nu}].
\label{EQ1}
\end{equation}
Here e has one mass dimension. The indices $\mu$ and $\nu$ takes
the values $0$ and $1$ in $(1+1)$ dimensional space time. The
dilaton field $\Phi$ stands as a non dynamical back ground and its
only role in this model is to make the coupling constant a
position dependent one. Let us now define $g^2(x) = e^{2\Phi(x)}$.
Here as usual we will choose a particular dilaton background
motivated by the linear dilatonic vacuum of $(1+1)$ dimensional
gravity. Therefore, $\Phi(x) = -x^1$, where $x^1$ is space like
coordinate. The region $x^1 \to\+ \infty$, corresponds to exterior
space where the coupling $g^2(x)$ vanishes and the fermion will be
able to propagate freely. However, the region where $x^1 \to
-\infty$, the coupling constant will diverge and it is analogous
to infinite throat in the interior of certain magnetically charged
black hole.

The equation (\ref{EQ1}) is obtained from the action
\begin{equation}
S_{AF} = \int d^2\sigma\sqrt{g}[R + 4(\nabla\phi)^2 + {1\over
{Q^2}} - {1\over 2}F^2 + i \bar\psi D\!\!\!/\psi] \label{EQ2}
\end{equation}
for sufficiently low energy incoming fermion and negligible
gravitational effect \cite{STRO}. It is a two dimensional
effective field theory of dilaton gravity coupled to fermion. Here
$\Phi$ represents the scalar dilaton field and $\psi$ is the
charged fermion. Equation (\ref{EQ2}) was derived viewing the
throat region of a four dimensional dilatonic black hole as a
compactification from four to two dimension \cite{GARF, GIDD,
STRO}. Note that, in the extremal limit, the geometry is
completely non-singular and there is no horizon but when a low
energy particle is thrown into the non-singular extremal black
hole, it produces a singularity and an event horizon. In this
context, we should mention that the geometry of the four
dimensional dilatonic black hole consists of three regions
\cite{GARF, GIDD, GIDD1, STRO}. First one is the asymptotically
flat region far from the black hole. As long as one proceed nearer
to the black hole the curvature begins to rise and finally enters
into the mouth region (the entry region to the throat). Well into
the throat region, the metric is approximated by the flat two
dimensional Minkosky space times the round metric on the two
sphere with radius Q and equation (\ref{EQ2}) results. The dilaton
field $\Phi$ indeed increases linearly with the proper distance
into the throat.

In the present situation we are interested in studying the
scattering of chiral fermion. So we need to replace the vector
interaction by the chiral interaction which leads to
\begin{equation}
{\cal S}_f = \int d^2x[i\bar\psi\gamma^\mu[\partial_\mu +
ieA_\mu(1+\gamma_5)]\psi - {1\over 4}
e^{-2\Phi(x)}F_{\mu\nu}F^{\mu\nu}]. \label{EQQ}
\end{equation}
Equation (\ref{EQQ}), is the quantum chiral electrodynamics in
place of quantum electrodynamics with a dilaton back ground field
$\Phi$. This can be decoupled into the following
\begin{eqnarray}
{\cal S}_f &=& \int d^2x[\bar\psi_R\gamma_\mu\partial^\mu\psi_R+
i\bar\psi_L\gamma^\mu(\partial_\mu + ieA_\mu)\psi_L \nonumber \\
&-& {1\over 4} e^{-2\Phi(x)}F_{\mu\nu}F^{\mu\nu}].
\end{eqnarray}
Here $\psi_R$ represents the right handed fermion. This right handed fermions
remain uncoupled in this type of chiral interaction.
Integration over the right handed part gives a field independent counter
part which can be absorbed within the normalization and the action
reduces to the following
\begin{equation}
{\cal S}_f
= \int d^2x[
i\bar\psi_L\gamma^\mu(\partial_\mu + i2e\sqrt{\pi}A_\mu)\psi_L \nonumber \\
- {1\over 4} e^{-2\Phi(x)}F_{\mu\nu}F^{\mu\nu}].\label{EQ5}
\end{equation}
In equation (\ref{EQ5}) e is replaced by $2e\sqrt{\pi}$ for later
convenience.
We now  bosonize the theory. The advantage of using the bosonized
version is that the anomaly automatically gets incorporated within
it. So the tree level bosonized theory contains the effect of
anomaly too. In order to bosonize the theory when we integrate out
the left handed fermion anomaly enter into the theory.  The
anomaly considered in this situation is of Faddeevian class
\cite{FADDEEV}.
 With the generalized
Faddeevian anomaly \cite{MG, AR1} the bosonized action reads
\begin{eqnarray}
{\cal L}_{CH} &=& {1\over 2}\partial_\mu\phi\partial^\mu\phi +
e(\eta^{\mu\nu} - \epsilon^{\mu\nu})
\partial_\mu\phi A_\nu \nonumber \\
&+& {1\over 2} e^2[A_0^2 - A_1^2 +2 \alpha A_1( A_0 + A_1)] -
{1\over 4} e^{-2\Phi(x)}F_{\mu\nu}F^{\mu\nu} .\end{eqnarray}
This
equation though shows no Lorentz covariant structure it has the
physical Lorentz invariance \cite{AR1}. Here we impose the chiral
constraint $\Omega(x) = \pi(x) - \phi'(x)$, to express the action
in terms of chiral boson following the procedure available in
\cite{KH}. In terms of chiral boson the model turns into
\begin{eqnarray}
{\cal L}_{CH} &=& \dot\phi\phi' -\phi'^2 + 2e(A_0 - A_1)\phi'\nonumber \\
&+& {1\over 2}e^2 [2(\alpha  - 1)A_1^2 + 2(\alpha + 1)A_0A_1] +
{1\over 2} e^{-2\Phi(x)}F_{01}^2.
\label {LCH}
\end{eqnarray}
Here $\phi$ represents a scalar field. Note that $\dot\phi^2$ is
absent because the first two terms in the lagrangian (\ref{LCH})
corresponds to the kinetic term for chiral boson \cite{SIG, FLO}.
It is now necessary to carry out the Hamiltonian analysis of the
theory to observe the role of dilaton field on the equation of
motion. From the standard definition of momentum the canonical
momenta corresponding to the chiral boson field $\phi$, the gauge
field $A_0$ and $A_1$ are obtained.
\begin{equation}
\pi_\phi = \phi',\label{MO1}
\end{equation}
\begin{equation}
\pi_0 = 0,\label{MO2}
\end{equation}
\begin{equation}
\pi_1 = e^{-2\phi(x)}(\dot A_1 - A_0')={1\over {g^2}}(\dot A_1 - A_0).\label{MO3}
\end{equation}
Here $\pi_\phi$, $\pi_0$ and $\pi_1$ are the momenta corresponding to the field $\phi$, $A_0$ and
$A_1$. Using the above equations
it is straightforward to obtain the canonical Hamiltonian through a Legendre
transformation. The canonical Hamiltonian is found out to be
\begin{eqnarray}
H_C &=& \int dx[{1\over 2} e^{2\Phi}\pi_1^2 + \pi_1A_0' + \phi'^2 -
 2e(A_0 - A_1)\phi' \nonumber \\
&-& {1\over 2}e^2[2(\alpha -1)A_1^2 + 2(1 + \alpha)A_0A_1)]].\label{CHAM}
\end{eqnarray}
The Hamiltonian though acquires an explicit space dependence
through the dilaton field $\Phi(x)$, it has no time dependence. So
it is preserved in time. Equation (\ref{MO1}) and (\ref{MO2}) are
the primary constraints of the theory. Therefore, it is necessary
to write down an  effective Hamiltonian:
\begin{equation}
H_{eff} = H_C + u\pi_0 + v(\pi_\phi - \phi'),
\end{equation}
where $u$ and $v$ are two arbitrary Lagrange multipliers. The
primary constraints (\ref{MO1}) and (\ref{MO2}) have to be
preserve in order to have a consistent theory. The preservation of
the constraint (\ref{MO2}), leads to a new constraint which is the
Gauss law of the theory:
\begin{equation}
G = \pi_1' + 2e\phi' +  e^2(1 + \alpha)A_1 = 0. \label{GAUS}
\end{equation}
The preservation of the constraint (\ref{MO1}) though does not give rise to
any new constraint it fixes the velocity $v$ which
comes out to be
\begin{equation}
v = \phi' - e(A_0 - A_1). \label{VEL}
\end{equation}
The constraint (\ref{GAUS}), also has to be conserved and the
conservation of it requires
\begin{equation} \dot G = 0\label{CON}
.\end{equation}
A new constraint
\begin{equation}
(1 + \alpha)e^{2\Phi}\pi_1 + 2\alpha (A_0' + A_1') = 0,\label{FINC}
\end{equation}
appears from the preservation condition (\ref{CON}).
No new constraints comes out from the preservation of (\ref{FINC}).
So we find that the phase space of the theory contains the following four constraints.
\begin{equation}
\omega_1 = \pi_0, \label{CON1}
\end{equation}
\begin{equation}
\omega_2 = \pi_1' + e\phi' +  e^2(1 + \alpha)A_1 = 0,\label{CON2}
\end{equation}
\begin{equation}
\omega_3 = (1 + \alpha)e^{2\Phi}\pi_1 + 2\alpha(A_0' + A_1') = 0,\label{CON3}
\end{equation}
\begin{equation}
\omega_4 = \pi_\phi - \phi'. \label{CON4}
\end{equation}
The four constraints (\ref{CON1}), (\ref{CON2}), (\ref{CON3}) and
(\ref{CON4}) form a second class set and all of these are weak
condition up to this stage. If we impose these constraints
strongly into the canonical Hamiltonian (\ref{CHAM}), the
canonical Hamiltonian gets simplified into the following.
\begin{equation}
H_R = \int dx^1[ {1\over 2}e^{2\Phi(x)}\pi_1^2 + {1\over {4e^2}}
\pi_1'^2 + {1\over 2}(\alpha - 1) \pi_1'A_1 + {1\over 4}e^2[(1 +
\alpha)^2 - 8\alpha)]A_1^2]. \label{RHAM}
\end{equation}
$H_R$ given in equation (\ref{RHAM}), is generally known as
reduced Hamiltonian. According to Dirac, Poisson bracket gets
invalidate for this reduced Hamiltonian \cite{DIR}. This reduced
Hamiltonian however remains consistent with the Dirac brackets
which is defined by
\begin{equation}
[A(x), B(y)]^* = [A(x), B(y)] - \int[A(x), \omega_i(\eta)] C^{-1}_{ij}(\eta, z)[\omega_j(z), B(y)]d\eta dz,
\label{DEFD}
\end{equation}
where $C^{-1}_{ij}(x,y)$ is defined by
\begin{equation}
\int C^{-1}_{ij}(x,z) [\omega_j(z), \omega_k(y)]dz =\delta(x-y) \delta_{ik}. \label{INV}
\end{equation}
For the theory under consideration
\noindent $C_{ij}(x,y) =$
\begin{equation}
 \pmatrix {0 & 0 & 2\alpha\delta'(x-y) & 0 \cr
0 & -2e^2(1+\alpha) \delta'(x-y) & e^2(1+\alpha)\delta(x-y) & 2e\delta'(x-y) \cr
& & -(\beta +1)\delta''(x-y)& \cr
2\alpha\delta'(x-y) & -e^2(1+\alpha)\delta(x-y)  & 2(\alpha+1)\times &  0 \cr
& +(\beta+1)\delta''(x-y) &(\beta+1)\delta'(x-y) & \cr
0 & 2e\delta'(x-y) & 0 & 2\delta'(x-y) \cr}.
\label{MAT}
\end{equation}
Here $i$ and $j$ runs from $1$ to $4$ and $\omega$'s represent the
constraints of the theory. With the definition (\ref{DEFD}), we
can compute the Dirac brackets between the fields describing the
reduced Hamiltonian $H_R$. The Dirac brackets between the fields
$A_1$ and $\pi_1$ are required to obtain the theoretical spectra
(equation of motion):
\begin{equation}
[A_1(x), A_1(y)]^* = {1\over {2e^2}}\delta'(x-y), \label{DR1}
\end{equation}
\begin{equation}
[A_1(x), \pi_1(y)]^* = {(\alpha -1)\over {2\alpha}}\delta(x-y),\label{DR2}
\end{equation}
\begin{equation}
[\pi_1(x), \pi_1(y)]^* = -{(1+\alpha)^2 \over {4\alpha}}e^2\epsilon(x-y).\label{DR3}
\end{equation}
Using (\ref{RHAM}), (\ref{DR1}), (\ref{DR2}) and (\ref{DR3}), we obtain the
following first order equations of motion:
\begin{equation}
\dot A_1 = {(\alpha-1) \over {2\alpha}} e^{2\Phi(x)}\pi_1 - A_1', \label{EQM1}
\end{equation}
\begin{equation}
\dot \pi_1 = \pi_1' + 2(\alpha - 1)e^2A_1. \label{EQM2}
\end{equation}
After a little algebra we find that the field $\pi_1$ satisfy a Klein Gordon
equation
\begin{equation}
(\Box - e^{2\Phi}e^2{(\alpha-1)^2\over {\alpha}})\pi_1 = 0 \label{SPEC},
\end{equation}
The equation (\ref{SPEC}), represents a massive boson with square
of the mass $m^2 = g^2{{-(1-\alpha)^2}\over \alpha }e^2$. Here
$\alpha$ must be negative in order to have the mass of the boson a
physical one. Mass of this boson however in this particular
situation is not constant. It contains a position dependent factor
$g^2$ which increases indefinitely in the negative $x^1$
direction. Thus any finite energy contribution must be totally
reflected. Therefore, an observer at $x^1 \to \infty$ will recover
all information. To be more specific, mass will vanish near the
mouth (the entry region to the throat) but increases indefinitely
as one goes into the throat because of the variation of this space
dependent factor $g$. Since massless scalar is equivalent to
massless fermion in $(1+1)$ dimension, we can conclude that a
massless fermion proceeding into the black hole will not be able
to travel an arbitrarily long distance and will be reflected back
with a unit probability. So, there will be no information loss and
a unitary s-matrix can be constructed for this particular
scattering problem. This results reminds us the scattering of
Dirac fermion \cite{STRO, SUS}. Thus in the description of chiral
fermion where $U(1)$ anomaly has been taken into account with the
introduction of Faddeevian class of anomalous term is found to be
free from the {\it dangerous} information loss problem. It is the
anomaly which has brought this noble change in the scattering
result and has saved this model from the danger of information loss.
Note that this result is just opposite to the conclusion of \cite{MIT}.
The analysis available
in \cite{MIT}, and the present analysis differs only in the anomaly structurebut this particular class of anomaly changes the preservation of
information scenario for chiral fermion significantly. It is
tempting to note that this noble change appeared in the scattering
of chiral fermion in connection with information loss scenario is
consistent with the standard belief as well as with the Hawking's
recent suggestion.
This has not come as a great surprise - similar situations
were found elsewhere in quantum electro dynamics and quantum
chiral electrodynamics \cite{JR, KH, PM, MG, AR1, AR}. A famous
instance is the removal of the long suffering of the chiral
electro dynamics from the non unitarity problem \cite{JR}. In
(1+1) dimensional field theories, we have also noticed the crucial
role of anomaly in the confinement aspect of fermion \cite{PM,
AR1, SCH, AR}.
In the present letter what we observed is a crucial but very
interesting role of anomaly. Here we observe its novel role in
connection with information loss paradox. So anomaly not always
appears as a disturbing term some times it appears as a surprise
to give a relief from disturbance. Of course, the merit of a 
particular model establishes itself when it leads to a definite 
positive conclusion.
This is not a special feature of chiral fermion only. One can also find the variation of 
information loss scenario in the Dirac fermion 
if we use aomalous  \cite{AR1} Schwinger
model, in place of vector Schwinger model and study the same scattering problem. The bosonised version of the anomalous Schwinger model coupled with the dilaton field is given by the lagrangian density
\begin{equation}
{\cal L} = {1\over 2}\partial_\mu\phi\partial^\mu\phi + \epsilon_{\mu\nu}
\partial^\nu\phi A^\mu + {1\over 2}ae^2A_\mu A^\mu + {1\over 4}e^{-2\Phi(x)}
F_{\mu\nu}F^{\mu\nu}
\end{equation}
If we proceed through the constraint analysis and obtain hamiltonian equations of motion using Dirac's prescription for constrained hamiltonian systems \cite{AR} we will land  into  the
 following two second order equations of motion.
\begin{equation}
(\Box+m^2)A_1=0,
\end{equation}
\begin{equation}
\Box\phi=0.
\end{equation}
Here $A_1$ represents a massive boson and $\phi$ represents a scalar field. Square of the mass $m^2$ takes the value $m^2=(a+1)e^2e^{2\Phi(x)}$. 
Detail calculation is not given here. $\Box\phi=0$, represents a massless boson which is equivalent to a massless fermion. This fermion will travel
within the black hole without any hindrance and information loss stood
as a real problem in this situation. It is to be noted that this problem was not there when anomaly was not taken into consideration, i.e., when a
gauge invariant bosonised lagrangian is used in the analysis \cite{STRO}.  We should mention here that gauge current in the present situation is 
\begin{equation}
J_\mu = e\epsilon_{\mu\nu}\partial^\nu\phi + ae^2A_\mu 
\end{equation} and it does not conserve. In the analysis available in \cite{STRO}, it was $J_\mu = e\epsilon_{\mu\nu}\partial^\nu\phi$ and 
needless to mention that it was conserved. 
Variation of information loss problem though looks strange nevertheless it is there within the fermion scattering problem because of the presence anomaly in the in the current and there is room for alteration of anomaly structure \cite{MIT, JR, KH, PM, MG, AR1}.
However one may raise the question which 
result is consistent in connection with information loss issue.
Obviously there is no specific answer. What actually happened in the
plank scale physics is not yet clearly known. It still remains as an unsetteled issue whether information loss
or information preserving result would be acceptable for fermion. If information
preserving result is considered to be the accepted one then in that case Faddeevian
anomaly will get a special status and our analyses correspocding to chiral fermion will score over the other. 

\noindent {\bf Acknowledgment}: It is a pleasure to thank the
Director, Saha Institute of Nuclear Physics and the Head of the
Theory Group of Saha Institute of Nuclear Physics, Kolkata for
providing working facilities.

\end{document}